\documentclass[11pt,reqno]{amsart}
\usepackage{latexsym}
\usepackage{amssymb}
\usepackage{amsmath}
\usepackage{amsthm}
\usepackage{amsfonts}

\usepackage{graphicx,color}  
\usepackage{graphpap}  
\usepackage{epic, xypic} 

\usepackage{mathrsfs, euscript}

\newcommand{\scro}{ {\mathcal O}}

\newcommand{\bp}{{\mathbb P^2}}
\newcommand{\scrop}{{\mathcal{O}}_{\bp}}

\usepackage{color}

\numberwithin{equation}{section}

\newcommand{\CC}{{\mb{C}}}

\newcommand{\PP}{{\mb{P}}}
\newcommand{\QQ}{{\mb{Q}}}

\newcommand{\ZZ}{{\mb{Z}}}

\newcommand{\ch}{{\rm ch}}
\newcommand{\td}{{\rm td}}

\newcommand{\eqn}{\begin{eqnarray}}
\newcommand{\feqn}{\end{eqnarray}}

\newtheorem{prop}{Proposition}

\ifx\mb\undefined
\newcommand{\mb}[1]{{\mathbb #1}}
\fi

\usepackage[colorlinks=true,          
            linkcolor=blue,
            citecolor=blue,
            urlcolor=blue,
            letterpaper,              
            bookmarks=true,           
            bookmarksnumbered=true,   
            bookmarksopen=true,       
            bookmarksopenlevel=1,     
            breaklinks=true,          
            pdfpagelayout=TwoPageRight, 
            pdfstartview=FitV,        
            pdfstartpage=1,           
            pdfdisplaydoctitle=true   
]{hyperref}
\hypersetup{ pdfcreator  = {LaTeX with hyperref package},
             pdfproducer = {dvips + ps2pdf} }
\let\url\nolinkurl 

\begin{document}

\title[The Physical Mirror Equivalence for the Local $\PP^2$]{The Physical Mirror Equivalence for the Local $\PP^2$}

\author{Sergio Luigi Cacciatori}
\address{
Dipartimento di Scienze ed Alta Tecnologia, Universit\`a degli Studi dell'Insubria,
Via Valleggio 11, 22100 Como, Italy, and INFN, via Celoria 16, 20133 Milano, Italy}
\email{sergio.cacciatori@uninsubria.it}
\author{Marco Compagnoni}
\address{
Dipartimento di Matematica, Politecnico di Milano, Via Bonardi 9, 20133 Milano, Italy}
\email{marco.compagnoni@polimi.it}
\author{Stefano Guerra}
\address{
Department of Mathematics, Oregon State University, Corvallis OR}
\email{guerras@math.oregonstate.edu}


\begin{abstract}
In this paper we consider the total space of the canonical bundle of $\PP^2$ and we use a proposal by Hosono, together with results of Seidel and
Auroux-Katzarkov-Orlov, to deduce the physical mirror equivalence between $D^b_{\PP^2}(K_{\PP2})$ and the derived Fukaya category of its mirror which assigns the expected central charge to BPS states. By construction, our equivalence is compatible with the mirror map relating the complex and the K\"ahler moduli spaces and with the computation of Gromov--Witten invariants.
\end{abstract}

\maketitle

\section{Introduction}\noindent
In \cite{h} Hosono describes a way to define the mirror map at the level of K-theory/homology based on homological mirror symmetry. In this paper we analyze Hosono's mirror symmetry prescription for the total space of the canonical bundle of the projective plane.\\
Let $X_t$ be a Calabi--Yau manifold (where $t$ denotes the complexified K\"ahler moduli), $Y_x$ its mirror family (where $x$ denotes local coordinates for the complex moduli space) and $\Omega(Y_x)$ the corresponding holomorphic 3-form.
Mirror symmetry identifies the K\"ahler moduli of $X_t$ with the periods of  $\Omega(Y_x)$ over some 3--cycles of $Y_x$. These, as functions of the complex structure moduli, must satisfy a system of Picard--Fuchs equations. For toric Calabi-Yau manifolds these are the GKZ equations \cite{GKZ}. From this follows that a particular cohomology valued hypergeometric series $w(x)$ arises naturally providing a basis of solutions for the GKZ system \cite{HKTY1,HKTY2,HLY1,HLY2}. Hosono was able to recognize such series as a formula \hbox{identifying} the BPS states of the associated physical theory.
Let $E_1,\ldots,E_r$ a basis of $K(X_t)$ and denote by $\chi^{ij} = \bigl( \chi(E_i, E_j) \bigr)^{-1}$ the inverse of the Euler characteristic. In \cite{h} Hosono proposed that
\begin{equation}\label{hosonoconj}
w(x) = \sum_{ij} \Bigl( \int_{S_i} \Omega(Y_x)\Bigr)\, \chi^{ij}\, ch(E_j^{\vee}),
\end{equation}
where $S_i$ in $H_3(Y_x,\QQ)$ is the homological mirror cycle of the class $\chi^{ij} E_j$ in $K(X_t)$. In order to clarify the physical meaning of the Hosono's conjecture, let $B_1,\ldots,B_r$ be a basis of the compactly supported K-theory $K^c(X_t)$ corresponding to a basis of admissible brane configurations, and $E_1,\ldots,E_r$ its dual basis in $K(X_t)$ w.r.t. the natural pairing between $K(X_t)$ and $K^c(X_t)$. Taking the basis $Q_i:=\ch(E_i),\ i=1,\ldots,r$ of $H^*(X_t,\QQ)$ and expanding  $w\left(x\right)=\sum_{i=1}^r w_i(x) Q_i$, the conjecture can be stated as follows:
\begin{enumerate}
\item\label{primo}
the coefficients $w_i(x)$ of the hypergeometric series may be identified with the period integrals over the mirror cycles $mir(B_i)$
\begin{eqnarray*}
w_i(x)=\int_{mir(B_i)} \Omega(Y_x)\ ;
\end{eqnarray*}
\item\label{secondo}
the monodromy of the hypergeometric series is integral and symplectic with respect to the symplectic form defined in $K^c(X_t)$
\begin{eqnarray*}
\chi(B_i,B_j)= \int_{X_t} \ch(B_i^\vee) \ch(B_j) \td(X_t)\ ;
\end{eqnarray*}
\item\label{terzo}
the central charge of an element $F\in K^c(X_t)$ is expressed in terms of the cohomology valued hypergeometric $w(x)$ as
\begin{eqnarray*}
Z(F)=\int_X \ch(F) w\left(x\right) \td(X_t)\ .
\end{eqnarray*}
\end{enumerate}
Therefore, through Hosono's proposal one has a linear map \footnote{This should actually be a symplectomorphism if we take as symplectic form in $H_3(Y_x, \QQ)$ the intersection form and in $K^c(X_t)$ the Euler characteristic.}
\begin{equation}
mir: K^{\rm c}(X_t) {\longrightarrow} H_3 (Y_x,\mathbb{Q}),
\end{equation}
that Hosono conjectured should induce an equivalence
\begin{equation}
Mir: D^{b} \bigl(Coh(X_t) \bigr) \longrightarrow D Fuk^o(Y_x,\omega).
\end{equation}
We will refer to this equivalence as the physical mirror map.
By construction, both $mir$ and $Mir$ are compatible with the mirror map between K\"ahler and complex moduli spaces, in particular $mir$ provides a direct link between the periods of the dual cycles and the Gromow-Witten invariants in the Gopakumar-Vafa interpretation. In this respect, the physical mirror map, among all the other autoequivalences, seems to have a privileged position also from the mathematical point of view. The Hosono's proposal has been verified in several examples in relation to the computation of the Gromov--Witten invariants (see \cite{h} and \cite{Cacciatori:2008fq} for the multiple resolution case), but until now there was no a direct check at the homological level.\\
Homological mirror symmetry for a del Pezzo surface, $S$, was considered in \cite{ako}. The authors found an equivalence between the derived
category of coherent sheaves of a del Pezzo surface and the derived Fukaya category of its mirror. This is achieved by matching the objects of an exceptional
sequence of sheaves in $D^b(Coh(S))$ with an exceptional sequence of Lagrangian cycles in the derived Fukaya category of the mirror, and checking that the
corresponding morphisms match as well. In \cite{Seidel1} Seidel considers homological mirror symmetry for the total space of the canonical bundle of a toric
del Pezzo surface $S$. If we denote by $M$ the mirror of $S$, and by $Y$ the mirror of $K_S$, Seidel shows how to lift the equivalence $D^b(Coh(S)) \rightarrow D Fuk^o(M)$ to
a full embedding $D^b_S(Coh(K_S)) \hookrightarrow D Fuk^o(Y)$ \footnote{Here $D^b_S(Coh(K_S))$ denotes the category of complexes of coherent sheaves on $K_S$ with cohomology supported on the zero section. }. The main tool introduced in
\cite{Seidel1} is the suspension of A$_{\infty}$-algebras.\\
In this manuscript, we introduce a new construction for lifting the Lagrangian cycles from the mirror of $\PP^2$ to the mirror of the total space of its canonical bundle which is equivalent to Seidel's suspension. Using these cycles we are able to compute the periods of the holomorphic 3-form $\Omega(Y_x)$ and using Hosono's observation (\ref{hosonoconj}) and Seidel's theorem \cite{Seidel1} we provide a full embedding of the derived category of coherent
sheaves of $K_{\PP^2}$ (with cohomology supported on the zero section) in the derived Fukaya category of its mirror. This full embedding is conjecturally an equivalence \cite{Seidel1}. As mentioned earlier, by construction, our correspondence induces the correct (physical) isomorphism at the level of K-theory/homology which computes the expected brane charges and it is compatible with the mirror map at the level of moduli spaces which is used in the computation of GW invariants. In this respect, we observe that it is necessary to twist by $\scro(-2)$ the equivalence found in \cite{ako} in order to induce the correct map at the level of K-theory/homology.\\
The paper is organized as follows: in section $2$ we recall few facts we need about the total space of the canonical bundle of $\PP^2$ and its mirror. We
introduce Hosono's construction \cite{h} and summarize the results from \cite{ako} and \cite{Seidel1} which we will use later in the paper. In section $3$ we consider
the mirror family $Y_x$ of $K_{\PP^2}$. We introduce a construction to lift the Lagrangian cycles from the mirror of $\PP^2$ to $Y_x$ which is
equivalent to Seidel's suspension. Next we compute the periods and use the result to derive the embedding $D^b_{\PP^2}(Coh K_{\PP^2}) \hookrightarrow D Fuk^o(Y)$.
In the last section we discuss our results. We collect in an appendix some of the facts and computations about hypergeometric functions that we use in
order to compute the periods in section $3$.

\section{Homological Mirror Symmetry for $K_{\PP^2}$}\noindent
In this section we summarize what we need about local homological mirror symmetry for $K_{\PP^2}$, which is the unique crepant resolution of the toric
orbifold $\CC^3/\ZZ_3$. In \cite{h} Hosono proposed a method to define explicitly the (physical) homological mirror map in the case of a non compact toric
Calabi--Yau threefolds. In \cite{Seidel1} Seidel gives a partial proof for the homological mirror symmetry conjecture in the case of the total space of the
canonical bundle $K_S$ of a toric del Pezzo surface $S$, extending the results of \cite{ako} for del Pezzo surfaces. In section \ref{PhysMirr} we will combine
these results finding the explicit homological mirror map for (the resolution of) $\CC^3/\ZZ_3$ using Hosono's construction.

\subsection{The Toric Orbifold $\CC^3/\ZZ_3$, the Crepant Resolution and its Mirror Family}
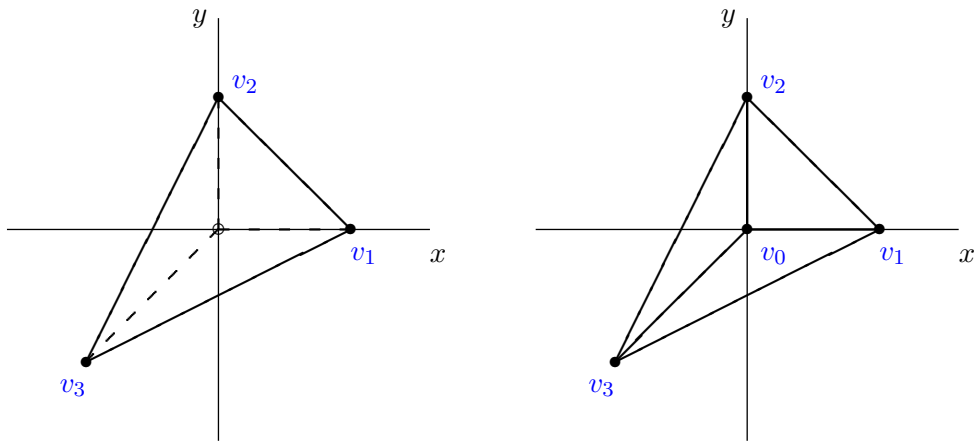
\begin{figure}[h]
\centering
\caption{Fan for ${\CC}^3/\ZZ_3$ (on the left) and its crepant resolution $X:=K_{\PP^2}$ (on the right).}\label{fig1}
\begin{picture}(0,190)(0,-90)
\put(-100,50){\circle*{4}}
\put(-100,0){\circle{4}}
\put(-50,0){\circle*{4}}
\put(-150,-50){\circle*{4}}
\drawline(-100,-80)(-100,80)
\drawline(-180,0)(-20,0)
\drawline(100,-80)(100,80)
\drawline(20,0)(180,0)
\thicklines
\dashline{5}(-100,0)(-50,0)
\dashline{5}(-100,0)(-100,50)
\dashline{5}(-100,0)(-150,-50)
\drawline(-150,-50)(-100,50)
\drawline(-150,-50)(-50,0)
\drawline(-50,0)(-100,50)
\put(-20,-10){\makebox(0,0)[l]{\color{black}{$x$}}}
\put(-110,80){\makebox(0,0)[l]{\color{black}{$y$}}}
\put(-50,-10){\makebox(0,0)[l]{\color{blue}{$v_1$}}}
\put(-95,55){\makebox(0,0)[l]{\color{blue}{$v_2$}}}
\put(-160,-60){\makebox(0,0)[l]{\color{blue}{$v_3$}}}
\put(100,50){\circle*{4}}
\put(100,0){\circle*{4}}
\put(150,0){\circle*{4}}
\put(50,-50){\circle*{4}}
\drawline(100,0)(150,0)
\drawline(100,0)(100,50)
\drawline(100,0)(50,-50)
\drawline(50,-50)(100,50)
\drawline(50,-50)(150,0)
\drawline(150,0)(100,50)
\put(180,-10){\makebox(0,0)[l]{\color{black}{$x$}}}
\put(90,80){\makebox(0,0)[l]{\color{black}{$y$}}}
\put(150,-10){\makebox(0,0)[l]{\color{blue}{$v_1$}}}
\put(105,55){\makebox(0,0)[l]{\color{blue}{$v_2$}}}
\put(40,-60){\makebox(0,0)[l]{\color{blue}{$v_3$}}}
\put(105,-10){\makebox(0,0)[l]{\color{blue}{$v_0$}}}
\end{picture}
\end{figure}

Let us consider the three dimensional complex orbifold
\begin{eqnarray}
\CC^3/\ZZ_3=\left\{(x,y,z)\in\CC^3|(x,y,z)\sim (\omega x,\omega y, \omega z), \ \omega^3=1\right\}.
\end{eqnarray}
It is a non compact toric variety\footnote{We refer to \cite{FultonT} for the details on toric geometry.} with fan $\Delta$ generated by the three vectors
\begin{eqnarray}
v_1=\begin{pmatrix}0\\1\\1 \end{pmatrix}, \qquad v_2=\begin{pmatrix}1\\0\\1 \end{pmatrix}, \qquad
v_3=\begin{pmatrix}-1\\-1\\1 \end{pmatrix}.
\end{eqnarray}
All relevant toric data are encoded in the 2-dimensional intersection of $\Delta$ with the plane $z = 1$ (F{\sc igure} \ref{fig1} left).

\noindent The space $\CC^3/\ZZ_3$ is a singular Calabi--Yau threefold and it is one of the most intensively studied examples in the context of local mirror
symmetry (see for example \cite{Chiang,ossa,Karp:2006pb,Bouchard:2007nr}). It has an isolated singularity at the origin and it admits unique crepant resolution.
The fan of this resolution is obtained from the one of $\CC^3/\ZZ_3$ by adding the vector
\begin{eqnarray}
v_0=\begin{pmatrix} 0\\0\\1 \end{pmatrix}
\end{eqnarray}
and the related faces (F{\sc igure} \ref{fig1} right). The resolved toric variety $X$ is the total space $K_{\PP^2}$ of the canonical bundle
${\mathcal O}_{\PP^2} (-3)$ of the projective plane.

We define the mirror of $X$ using the local generalization of the Batyrev construction for noncompact toric varieties \cite{Chiang}.  From the toric data of
$X$ we define the superpotential:
\begin{eqnarray}
W(x_1,x_2;U,V;\vec a)=UV+a_0+a_1 x_1+ a_2 x_2 +a_3 x_1^{-1} x_2^{-1},
\end{eqnarray}
with variables $(x_1,x_2;U,V)\in(\CC^*)^2 \times \CC^2$ and parameters $(a_0,\ldots,a_3)\in(\CC^*)^4$.
Rescaling the coordinates, the parameters can be expressed in terms of the invariant combination $y=a_1 a_2 a_3/a_0^3$. Consider the family of hypersurfaces $Y_y$
in $(\CC^*)^2 \times \CC^2$, parameterized by  $y\in\CC$, defined by the equation:
\begin{eqnarray}\label{fibration}
W(x_1,x_2;U,V;y)=UV+1+x_1+ x_2 +y x_1^{-1} x_2^{-1}=0.
\end{eqnarray}
In \cite{Chiang} it is claimed that $Y_y$ is the mirror family of $X$, where the complex parameter $y$ is mirror dual to the complexified K\"ahler parameter of $X$.

\subsection{Hosono's Construction}
Let $X_{\mathbf t}$ be a Calabi--Yau variety, with K\"ahler moduli $\mathbf t$, and $Y_{\mathbf y}$ its mirror family, with complex moduli $\mathbf y$. According
to Kontsevich's homological mirror conjecture \cite{kontsevich} there is an equivalence of triangulated categories:
\begin{equation}
Mir: D^{b} \bigl(Coh(X_{\mathbf t}) \bigr) \longrightarrow D Fuk^o(Y_{\mathbf y},\omega).
\end{equation}
The bounded derived category of coherent sheaves $D^{b} \bigl(Coh(X_{\mathbf t}) \bigr)$ depends on the choice of a complex structure on $X$, mirror to a
(complexified) symplectic structure $\omega$ on $Y$ used to define the derived Fukaya category $ DFuk^o(Y_{\mathbf y},\omega)$. The functor $Mir$ induces a
linear transformation
\begin{equation}
mir: K^{\rm c}(X_{\mathbf t}) {\longrightarrow} H_3 (Y_{\mathbf y},\mathbb{Q}),
\end{equation}
which is symplectic with respect to the symplectic form on $K^{\rm c}(X_{\mathbf t})$  given by the Euler characteristic $\chi$ and the one on
$H_3 (Y_{\mathbf y},\mathbb{Q})$ given by the intersection form.

\noindent
For noncompact Calabi--Yau manifolds, Hosono's conjecture completely characterizes the linear map $mir$. Later we will refer to the mirror map constructed
following Hosono's prescription as the ``physical'' mirror map. When $X:= K_{\PP^2}$ the conjecture can be stated as follows. Consider an
integral\footnote{This is an integral basis of $K^{\rm c}(X)$ since it is related to the dual integral basis of $K(X)$ given by the tautological sheaves on $X$
\cite{Ito-Nakajima,Craw} via an integral change of basis.} (brane) basis of $K^{\rm c}(X)$:
\begin{eqnarray}
\pmb \omega=\{{ {\mathcal B}}_0,{ {\mathcal B}}_1,{ {\mathcal B}}_2\}=
\Bigl\{ [{ {\mathcal O}}_p],[{ {\mathcal O}}_H (-1)],[{ {\mathcal O}}_{\PP^2}(-2)] \Bigr\}
\end{eqnarray}
where the square brackets are used to denote the class in K-theory of a sheaf.
Using the perfect pairing
\begin{equation}
\chi : K^c(X)\times K(X)\rightarrow {\mathbb{Z}}
\end{equation}
and the Chern character isomorphism
\begin{equation}
ch : K(X) \rightarrow H^*(X,\QQ),
\end{equation}
one defines a (group) isomorphism
\begin{equation}
\phi : K^c(X)\rightarrow H^*(X,\QQ).
\end{equation}
The rational cohomology ring of $X$ is
\begin{equation}
H^*(X,\QQ)=\QQ[J]/J^3,
\end{equation}
where $J$ is the Poincar\'e dual of the class of a line $H$ in the base $\PP^2$ and it is the generator of the K\"ahler cone of $X$. The (brane charges) basis
$\pmb {\tilde\omega}=\phi(\pmb {\omega})$ of $H^*(X,\QQ)$ is:
\begin{eqnarray}
\pmb {\tilde\omega}=\{{ Q}_0,{ Q}_1,{ Q}_2\}=
\Bigl\{ ch({ {\mathcal O}}_X),
ch \bigl({ {\mathcal O}}_X(J)-{ {\mathcal O}}_X\bigr),
ch \bigl({ {\mathcal O}}_X(2J)-2\,{ {\mathcal O}}_X(J)+{ {\mathcal O}}_X\bigr)\Bigr\}.
\end{eqnarray}
Then, using the toric data of $X$, we define the cohomological valued hypergeometric function (CHF)
\begin{eqnarray}
\left. w(y;J):=\sum_{n=0}^\infty \frac {y^{n+\rho}}{\Gamma(1+n+\rho)^3 \Gamma (1-3(n+\rho))}\right|_{\rho=\frac {J}{2\pi i}}
\end{eqnarray}
and we expand it in terms of the brane charges:
\begin{eqnarray}\label{chf}
&& w(y;J)=w_0 (y)  Q_0 +w_1(y)  Q_1 +w_2(y)  Q_2,\\
&& w_0(y)=1,\\
&& w_1(y)= \frac 1{2\pi i}\ln (y)+ \frac 3{2\pi i}\sum_{m=1}^{\infty} \frac {(3m-1)!}{(m!)^3} (-y)^m,\\
&& w_2(y)=-\frac 1{8\pi^2} (\ln (-y))^2 +\frac 18 -\frac {3}{4\pi^2} \ln (-y)\sum_{m=1}^{\infty} \frac {(3m-1)!}{(m!)^3} (-y)^m\cr
&& \phantom{w_2(y)=}-\frac {9}{4\pi^2} \sum_{n=1}^{\infty} \frac {(3n-1)!}{(n!)^3} [\psi(3n)-\psi(n+1)] (-y)^n,
\end{eqnarray}
where the $w_i$'s are solutions of the Picard-Fuchs equation.
Hosono conjectured that the components $w_i(y)$ are exactly the periods of the holomorphic 3-form $ \Omega(Y_y)$ over the three cycles in $Y_y$, mirror to
${ {\mathcal B}}_i$ :
\begin{equation}
w_0(y)=\int_{ mir({ {\mathcal O}}_p)}  \Omega (Y_y),\quad
w_1(y)=\int_{ mir({ {\mathcal O}}_H)}  \Omega (Y_y),\quad
w_2(y)=\int_{ mir({ {\mathcal O}}_{\PP^2})}  \Omega (Y_y).
\label{mmh}
\end{equation}
In section \ref{PhysMirr} we will use Hosono's construction to define the mirror functor $Mir$. We will find that this functor will recover the lift of the
equivalence found in \cite{ako} after applying an autoequivalence of the derived category $D^b_{\PP^2} (Coh(K_{\PP^2}))$ (see section \ref{mirrormap} for details).

\subsection{The Mirror of $\PP^2$}
In this section we recall few facts about mirror symmetry for $\PP^2$ following \cite{ako}.
The mirror of $\PP^2$ is the Landau-Ginzburg (LG) model given by the superpotential $W: \CC^{*2} \to \CC$:
\begin{eqnarray}
W(x_1,x_2)=x_1+x_2+\frac y{x_1x_2}.
\end{eqnarray}
If we denote by $M$ the partial compactification of $\CC^{*2}$ along the fibers of $W$, we get an elliptic fibration  over $\mathbb C$, $M\rightarrow\CC$.
Introducing the coordinate $z$ on the base of this fibration, the mirror becomes the surface $Z$ defined by the equation $W(x_1,x_2)=z$:
\begin{eqnarray}
x_1 x_2^2+x_1^2x_2+y+zx_1x_2=0
\end{eqnarray}
There are $3$ nodal singular fibers over the points:
\begin{eqnarray}
z_i=3\omega^i y^{\frac 13}, \qquad i=0,1,2,\qquad \omega=e^{\frac {\pi i}3}.
\end{eqnarray}
Using Seidel's construction of the Fukaya category (see \cite{Seidel2}), one can prove that there is an equivalence between $D^b \bigl(Coh(\PP^2) \bigr)$ and
$D\ Fuk^o(Z,\omega)$.
The category $D^b \bigl(Coh(\PP^2)\bigr)$ is generated by the exceptional collection
\begin{eqnarray}
\label{gencoh}
\pmb {\tau}\equiv \left\{ {E}_0, {E}_1, {E}_2\right\}:=\left\{ {\mathcal O}, {\mathcal T}(-1),
\Lambda^2 {\mathcal T}(-2)\simeq  {\mathcal O}(1)\right\}.
\end{eqnarray}

\noindent Using Picard-Lefschetz theory, one can argue that the derived Fukaya category $D\, {\rm Fuk}^o(Z, \omega)$ of the mirror fibration is also generated
by an exceptional collection of Lagrangian thimbles. The mirror LG model is endowed with a symplectic form $\omega$, which defines a horizontal distribution on
the elliptic fibration $W: M \to \CC$. Given a reference point $z_*$ on the base $\CC$  of the fibration, $z_* \neq z_i$ for $i= 1, 2, 3$, let $\Sigma_*$ be the
(smooth) fiber at $z_*$. Then we consider three paths $\gamma_i$ from $z_*$ to  $z_i$, intersecting only at $z_*$.  For each path $\gamma_i$ we consider a
vanishing cycle $L_i$ in $\Sigma_*$ which collapses at $z_i$. The parallel transport of $L_i$ along $\gamma_i$ defines the thimble ${D}_i$, having boundary
$\partial{D}_i= L_i$. These are the Lagrangian thimbles, whose intersection form can be expressed in terms of the relative cohomology w.r.t. $\Sigma_{*}$ and
it is equivalent to the one associated to $H_1(\Sigma_{*},\ZZ)$. 
The derived Fukaya category $D Fuk^o(Z, \omega)$ is equivalent to the category of vanishing cycles $D^b (Lag_{vc}(W\{\gamma_i\}))$ generated by the exceptional
collection
\begin{eqnarray}
\label{genvan}
\pmb {\tau'}\equiv \{{L}_0,{L}_1,{L}_2\}
\end{eqnarray}
(or more precisely by Hamiltonian isotopy classes of the thimbles ${D}_i$). The morphisms are defined in terms of the Floer cohomology:
\begin{eqnarray}
{\rm Hom}(L_i,L_j)=\left\{ \begin{array}{ccc}
FC^* (L_i, L_j;\CC)=\CC^{| L_i\cap L_j|}=\CC^3 & {\rm if}  & i<j \\
\CC                                                      & {\rm if}  & i=j \\
0                                                         & {\rm if}  & i>j
\end{array} \right.\ .
\end{eqnarray}
with composition:
\begin{eqnarray}
m_k : {\rm Hom} ( L_{i_0}, L_{i_1}) \otimes \ldots \otimes {\rm Hom} ( L_{i_{k-1}}, L_{i_k})
\rightarrow {\rm Hom} ( L_{i_0}, L_{i_k})[2-k]
\end{eqnarray}
which is trivial unless $i_0<i_1<\ldots < i_k$.

\noindent
In \cite{ako} the authors consider the case of a del Pezzo surface and prove that the objects, the morphisms and the composition law of the exceptional
collections (\ref{gencoh}) and (\ref{genvan}) are match. This is sufficient to prove (half of) the homological mirror conjecture, and it suggests that an
equivalence of categories could be given by:
\begin{eqnarray}
\label{mirmapako}
\left({\mathcal O}_{\PP^2}, {\mathcal T}_{\PP^2}(-1), {\mathcal O}_{\PP^2}(1) \right)\mapsto \left([ L_0], [ L_1], [ L_2]\right).
\end{eqnarray}

\subsection{The Mirror of $K_{\PP^2}$ as a Double Suspended Lefschetz Fibration}
Let us consider a del Pezzo surface $S$ with mirror LG--model defined by a superpotential $W(x_1,x_2)$. The mirror of the total space of the canonical bundle of
$S$, $K_S$, can be defined as the hypersurface \cite{hiv}:
\begin{eqnarray}\label{mirrorofx}
  Y=\left\{ (x_1,x_2; y_1, y_2)\in \CC^{*2} \times \CC^2 | W(x_1,x_2)-y_1^2-y_2^2=0 \right\}.
\end{eqnarray}
which agrees with \ref{fibration} when $S$ is the projective plane.
The local homological mirror symmetry for $K_S$ has been recently investigated. The relation between the category $D^b (Coh(S))$ and the category
$D^b_S (Coh(K_S))$, whose objects are complexes of sheaves on $K_S$ with cohomology supported on $S$, was studied in \cite{Ballard} and \cite{Segal}. In
\cite{Seidel1} Seidel extended the results of \cite{ako} to toric del Pezzo surfaces, proving the existence of a full embedding of $D^b_S (Coh(K_S))$ into
the derived Fukaya category of Lagrangian 3-cycles in $Y$:
\begin{eqnarray}
 D^b_S (Coh(K_S)) \hookrightarrow {Lag}_{vc} (Y).
\end{eqnarray}
The image of $D^b_S (Coh(K_S))$ is defined through a double suspension of the Fukaya category of the mirror of $S$. Let us consider the suspension
$W\mapsto W^\sigma$ of the superpotential:
\begin{eqnarray}
W^\sigma(x_1,x_2; y_1):=W(x_1,x_2)-y_1^2.
\end{eqnarray}
Then $W(x_1,x_2)-y_1^2 = 0$ defines a double cover of $M$ branched over $W(x_1,x_2)= 0$, which is a smooth fiber of the elliptic fibration $M \to \CC$. Since
the singular fibers of $M \to \CC$ are not over $z = 0$, to their singular point correspond two distinct inverse images in $W(x_1,x_2)-y_1^2 = 0$. Therefore to
the thimble $D_i$ of $M$ corresponds a thimble, in $W(x_1,x_2)-y_1^2 = 0$, having boundary the 2-sphere ${\overline S}_i$ obtained by gluing two copies of the
thimble $D_i$ (more precisely the inverse image of $D_i$ under the double cover) along their common boundary $L_i$. The ${\overline S}_i$'s are the vanishing
cycle of $W^{\sigma}$ lifting the $L_i$'s. In the same way we can suspend $W^\sigma$ w.r.t. a second variable $y_2$, obtaining the superpotential
\begin{eqnarray}
W^{\sigma\sigma}(x_1,x_2; y_1, y_2) =W(x_1,x_2)-y_1^2-y_2^2
\end{eqnarray}
and the double suspension of $L_i$ is a 3-dimensional sphere in $Y$.
The double suspensions of the cycles $L_i$ define a set of generators of a full subcategory of ${Lag}_{vc} (Y)$, and conjecturally of ${Lag}_{vc} (Y)$. In this
way Seidel extended the mirror map ($\ref{mirmapako}$) in the $K_S$ case, for $S$ toric.\\
In the next section we will show how our construction is related to the Seidel's double suspension and we will check the compatibility of Hosono's conjecture with the construction in \cite{ako} .

\section{Explicit Construction of the Mirror Functor.}
\label{PhysMirr}
Following Hosono's construction in \cite{h}, one can define a functor $D^{b}_{ {\mathbb{P}}^2}\bigl(Coh(X) \bigr) \rightarrow {Lag}_{vc}(Y)$ by sending a set
of generators of $D^b_{\PP^2} \bigl(Coh(X)\bigr)$ to a set of generators of the category ${Lag}_{vc}(Y)$. Our strategy for determining this functor is the
following. We will construct a set of Lagrangian 3-cycles generating
(a full subcategory of) ${Lag}_{vc}(Y)$ and will compute the corresponding integrals of the holomorphic 3-form $\Omega_y$. These will be
well-defined functions $F_i(y)$ of the complex modulus $y$. Under the mirror map, these functions must coincide with the coefficients of the cohomological
hypergeometric
function expanded in the basis of sheaves mirror to the Lagrangian cycles. In (\ref{chf}) we computed the functions corresponding to a selected basis of
D-branes in $X$. The mirror basis is obtained from this one by a linear transformation which will be determined by comparing the coefficients.

\subsection{The Double Fibration}
The mirror (\ref{mirrorofx}) of $X = K_{\PP^2}$ is defined by the equations:
\begin{equation*}
x_1 x_2^2+x_1^2x_2+y+zx_1x_2=0 \qquad UV + z = 0.
\end{equation*}
The first equation can be thought as an elliptic fibration over $\CC$ (parametrized by $z$), and the second equation as a conic fibration over the same base. The
mirror of $X$ is then the fiber product of these two fibrations. We will refer to this structure as ``double fibration'' and to the fibration associated to
$UV= -z$ as ``$\CC^*$-fibration''.\\
In order to construct a suitable set of Lagrangian 3-cycles, it is better to rewrite the double fibration above in a more standard form.
Following \cite{hiv}, we can rewrite the elliptic fibration in an almost Weierstrass form as follows. After rescaling the
coordinates by $x_i \to y^{\frac 13} x_i$ and taking as coordinate on the base of the fibration:
\begin{eqnarray*}
z=-\frac {UV}{y^{\frac 13}}-\frac 1{y^{\frac 13}}
\end{eqnarray*}
the equation for the elliptic fibration is
\begin{eqnarray}
x_1^2 x_2+x_2^2 x_1 -zx_1x_2+1=0.
\end{eqnarray}
Introducing the homogeneous coordinates
\begin{eqnarray*}
(X_0:X_1:X_2)=(1:x_1:x_2),
\end{eqnarray*}
it takes the form
\begin{eqnarray}
X_1^2 X_2+X_2^2 X_1 -zX_1X_2 X_0+X_0^3=0.
\end{eqnarray}
The holomorphic form now is
\begin{eqnarray}
\Omega_3=  \frac 1{(2\pi i)^3}\left(\frac {dX_1}{X_1} \wedge \frac {dX_2}{X_2} -\frac {dX_0}{X_0} \wedge \frac {dX_2}{X_2}
+\frac {dX_0}{X_0} \wedge \frac {dX_1}{X_1} \right) \wedge \frac {du}{u}. \label{holom3}
\end{eqnarray}
Next, we pass to the new coordinates $(X,Y,U)$ defined by
\begin{eqnarray*}
&& X_0=X, \qquad X_1=Y+\frac U2 +\frac {Xz}2, \qquad X_2=-Y+\frac U2 +\frac {Xz}2, \label{change}
\end{eqnarray*}
and restrict to the chart $U \neq 0$. Then, the elliptic fibration becomes
\begin{eqnarray}
&& Y^2 =X^3 +\left(\frac z2 \right)^2 X^2 +\frac z2 X+\frac 14. \label{cubic}
\end{eqnarray}
In particular the critical values of the elliptic fibration are
\begin{eqnarray}
z_0=3, \qquad z_1=3\omega, \qquad z_2=3\omega^2,
\end{eqnarray}
where $\omega:= -\frac 12 - i \frac {\sqrt 3}2$, whereas the singular point for the $\CC^*$ fibration is
\begin{eqnarray}
z_*=-\frac 1{y^{\frac 13}}.
\end{eqnarray}

\subsection{The Lagrangian Cycles}
\begin{figure}[h!]
\centering
\includegraphics[scale=0.7]{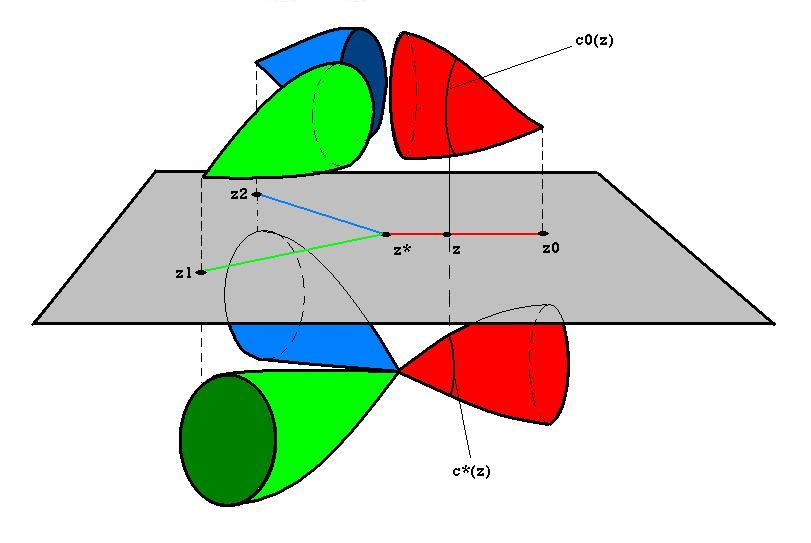}
\caption{A basis of vanishing Lagrangian 3-cycles.}\label{cicli-lag}
\end{figure}

The Lagrangian cycles of the mirror manifold $Y$ can be realized using the double fibration structure of $Y$ we have just described.
This fibration is singular at four points:
$z_*$, where the ${\CC^*}$-fiber becomes singular, and $z_i$, $i=0,1,2$, where the elliptic fibration becomes singular. Given a symplectic form both on the
elliptic and $\CC^*$-fibration one can define a symplectic connection and vanishing cycles. We choose cycles on the $\CC^*$-fibers over the points $z_i$ which
vanish at $z_*$ and on the elliptic fiber over $z_*$  which
vanish in $z_i$. By fixing a path  from $z_*$ to $z_i$ on the base of the double fibration, $\CC$, and taking the fiber product (over this path) of the $\CC^*$
cycle vanishing at $z_*$, and the cycle $c_i$ of the elliptic fiber vanishing at $z_i$, we get a 3-cycle diffeomorphic to a 3-sphere $S^3$ (F{\sc igure}
\ref{cicli-lag}).

\begin{figure}[h!]
\centering
\includegraphics[scale=0.45]{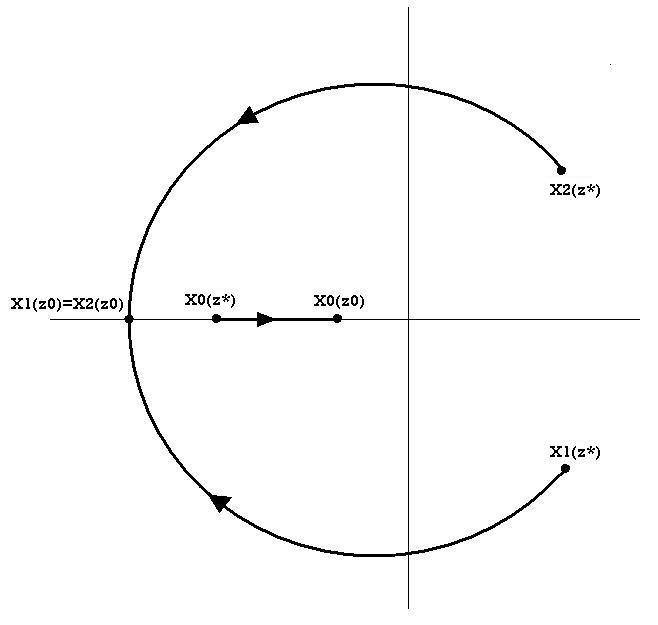}
\caption{Behaviour of the solutions $X_i(z)$ when the base point moves from $z=z_*$ to the critical point $z=z_0$.}\label{collapse}
\end{figure}

\begin{figure}[h!]
\centering
\includegraphics[scale=0.45]{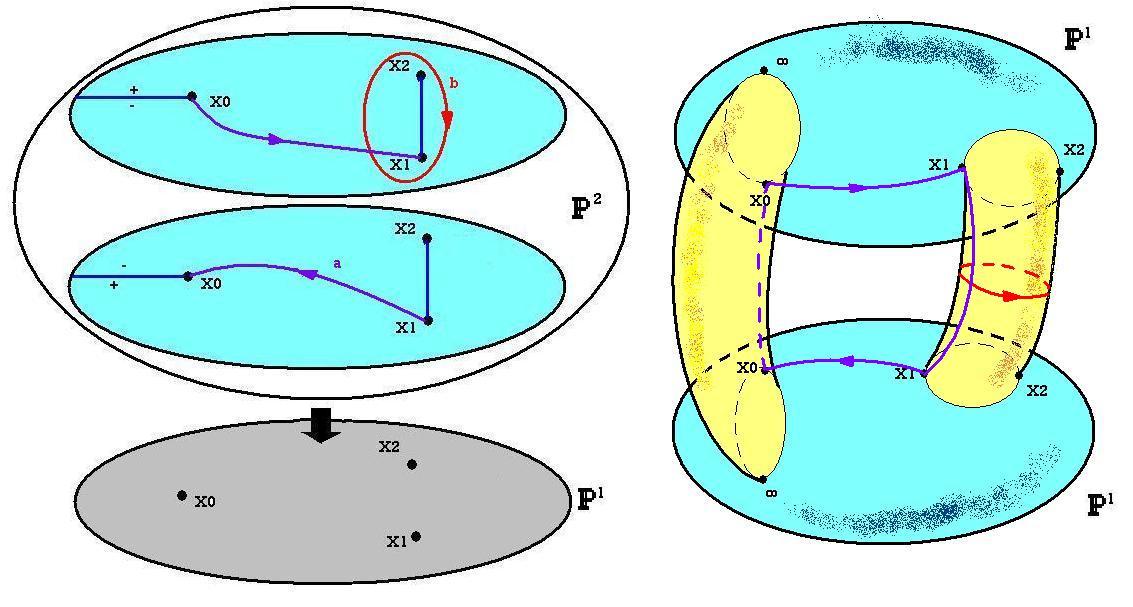}
\caption{Elliptic curve in ${\mathbb P}^2$ as a double cover of ${\mathbb P}^1$. The branch points are at $X_i$, $i=0,1,2$ and at $\infty$
and the gluing of the leaves is along two cuts, one from $X_1$ to $X_2$ and the other from $X_0$ to $\infty$. A canonical basis is given by the curve $a$ going
from $X_0$ to $X_1$ in the upper leaf and back from $X_1$ to $X_0$ in the lower leaf, and the curve $b$ encircling $X_1$ and $X_2$ in the
upper leaf.}\label{canonical-basis}
\end{figure}

\noindent In order to compute the periods we need to be more explicit. The zeros of to the cubic polynomial in (\ref{cubic}) are
\begin{eqnarray}
X_j(z)=\omega^{j} \left[ -\frac 1{27} \left( \frac z2\right)^6 +\frac 16 \left( \frac z2\right)^3-\frac 18 +
\frac 18 \sqrt {1-\left( \frac z3\right)^3} \right]^{\frac 13} + \nonumber \\
+\omega^{2j} \frac z6
\frac {\frac 13 \left( \frac z2\right)^3-1}{\left[ -\frac 1{27} \left( \frac z2\right)^6 +\frac 16 \left( \frac z2\right)^3-\frac 18 +
\frac 18 \sqrt {1-\left( \frac z3\right)^3} \right]^{\frac 13}}
-\frac 13 \left( \frac z2 \right)^2
\end{eqnarray}
for $ j=0,1,2$.
At the critical points we get
\begin{eqnarray}
&& \vec X(z_0)=\frac 14
\left(\begin{array}{r}
2\\-1\\-1
\end{array} \right)
-\frac 13 \left( \frac {z_0}2 \right)^2 \begin{pmatrix} 1\\1\\1 \end{pmatrix}, \\
&& \vec X(z_1)=\frac {\omega^2}4
\left(\begin{array}{r}
-1\\2\\-1
\end{array} \right)
-\frac 13 \left( \frac {z_1}2 \right)^2 \begin{pmatrix} 1\\1\\1 \end{pmatrix}, \\
&& \vec X(z_2)=\frac \omega4
\left(\begin{array}{r}
-1\\-1\\2
\end{array} \right)
-\frac 13 \left( \frac {z_2}2 \right)^2 \begin{pmatrix} 1\\1\\1 \end{pmatrix}.
\end{eqnarray}
Moreover, for the computations which we will carry out in the next section it is also useful to consider the values of $X_j(z)$ for $z = 0$
\begin{eqnarray}
X_j(0)=-2^{-\frac 23} \omega^j,\quad j=0,1,2. \label{radici}
\end{eqnarray}
\noindent At $z_*=-y^{-\frac 13}$ the roots are distinct but the vanishing cycle of the $\CC^*$ fibration shrinks to a point.
On the other hand, for $z = z_i$ we see that $X_j(z_i)=X_k(z_i)$ if $j\neq i\neq k$.\\
To determine the vanishing cycles we follow a procedure similar to the one used in \cite{ako} and \cite{ako1}. Let $\Sigma_*$ be the smooth fiber of the elliptic
fibration $M \to \CC$ over the reference point $z_*$. Consider the double cover $\Sigma_* \to \PP^1$, then the vanishing cycles $c_i$ are the double covers of
paths in $\PP^1$. The path in $\PP^1$ corresponding to $c_0$ is represented in F{\sc igure} \ref{collapse}. One can draw in a similar way the paths corresponding
to $c_1$ and $c_2$.\\
All smooth elliptic fibers near $\Sigma_*$ are diffeomorphic to $\Sigma_*$ and in what follows we will implicitly make this identification. Then, the vanishing
cycles $c_i(z)$, as $z$ varies along a path from $z_*$ to $z_i$, can be thought as a family of loops in $\Sigma_*$ whose projection to $\PP^1$ is drawn in
F{\sc igure} \ref{collapse} (for $c_0$). As $z$ varies between $z_*$ and $z_i$, the image of the vanishing cycle in $\PP^1$ is a path where the end points
move toward each other along the path between $x_2$ and $x_1$ and eventually collapse into each other on the negative real axis.\\
As mentioned above for $z$ varying between $z_*$ and $z_i$ we also have vanishing cycles on the $\CC^*$ fibration. These all collapse for $z= z_*$. We will
abuse notation and denote all three of them by $c_*$.  The 3-cycles we will consider are given by $c_i(z) \times c_*(z)$ for $i = 1, 2, 3$ and for $z$ varying
on a straight line joining $z_*$ to $z_i$ (F{\sc igure} \ref{cicli-lag}). When $z$ is not one of the critical points, $c_i(z)\times c_*(z) \cong T^2_i(z)$ is
diffeomorphic to a 2-torus, therefore, along a straight line joining $z_*$ to $z_i$, we get a torus fibration where the cycle $c_*(z)$ collapses at $z_*$ and
$c_i(z_i)$ collapses at $z_i$. More precisely we can choose $\Gamma_i:=\{z_i(t)=z_*+(z_i-z_*)t|t\in[0,1]\}$, and deform $c_i(z_i(t))$ homotopically to a circle
having radius $Rt$ and
$c_*(z_i(t))$ to a circle having radius $R\sqrt{1-t^2}$: $c_i(z_i(t))=\{ Rt e^{i\phi}|\phi\in [0,2\pi] \}$,
$c_*(z_i(t))=\{ R\sqrt {1-t^2} e^{i\psi}|\psi\in [0,2\pi] \}$. The closed cycle is thus equivalent to
\begin{eqnarray}
\begin{array}{ll}
{C}_i & \simeq \left\{(z,w)\in \CC^2| z=Rt e^{i\phi}, w=R\sqrt {1-t^2} e^{i\phi}, \phi\in [0,2\pi], \psi\in [0,2\pi], t\in [0,1] \right\} \\
\\
\phantom{\pmb{C}_i} & =  \left\{(z,w)\in \CC^2| |z|^2+|w|^2 =R^2\right\}\simeq S^3.
\end{array}
\end{eqnarray}
In F{\sc igure} \ref{canonical-basis} we give a representation of the elliptic curve $\Sigma_*$ with a choice
of a canonical basis for $H_1$, $\{a, b\}$.

\begin{figure}[h!]
\centering
\includegraphics[scale=0.45]{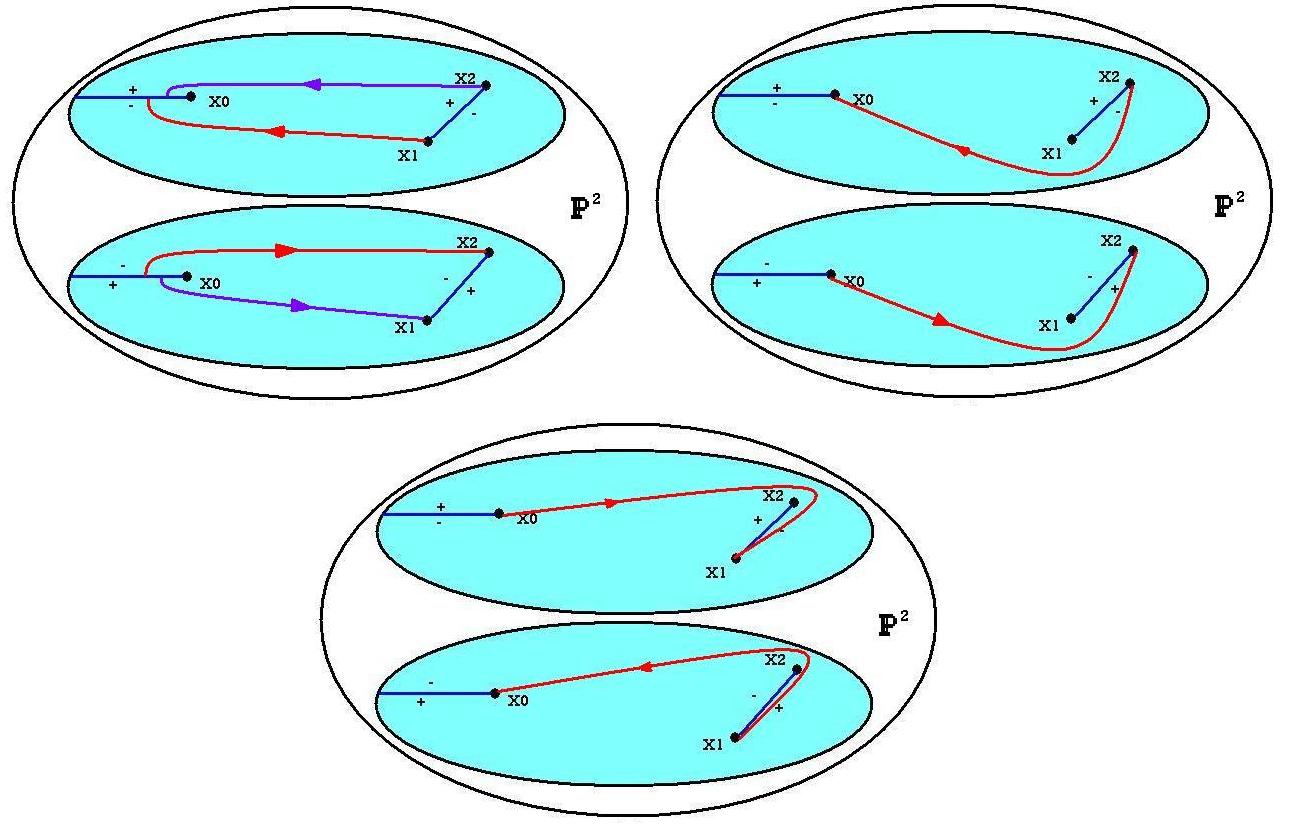}
\caption{Starting from the left, in clockwise order, the curves $c_0, c_1$ and $c_2$.}\label{c-curve}
\end{figure}

Our choice of the cycles $c_i$ in $\Sigma_*$ is given in F{\sc igure} \ref{c-curve}, and one can see that $c_0$ is homotopic to $-2a-b$, $c_1$ is
homotopic to $b- a$ and $c_2$ to $a+ 2b$.

\begin{figure}[h!]
\centering
\includegraphics[scale=0.45]{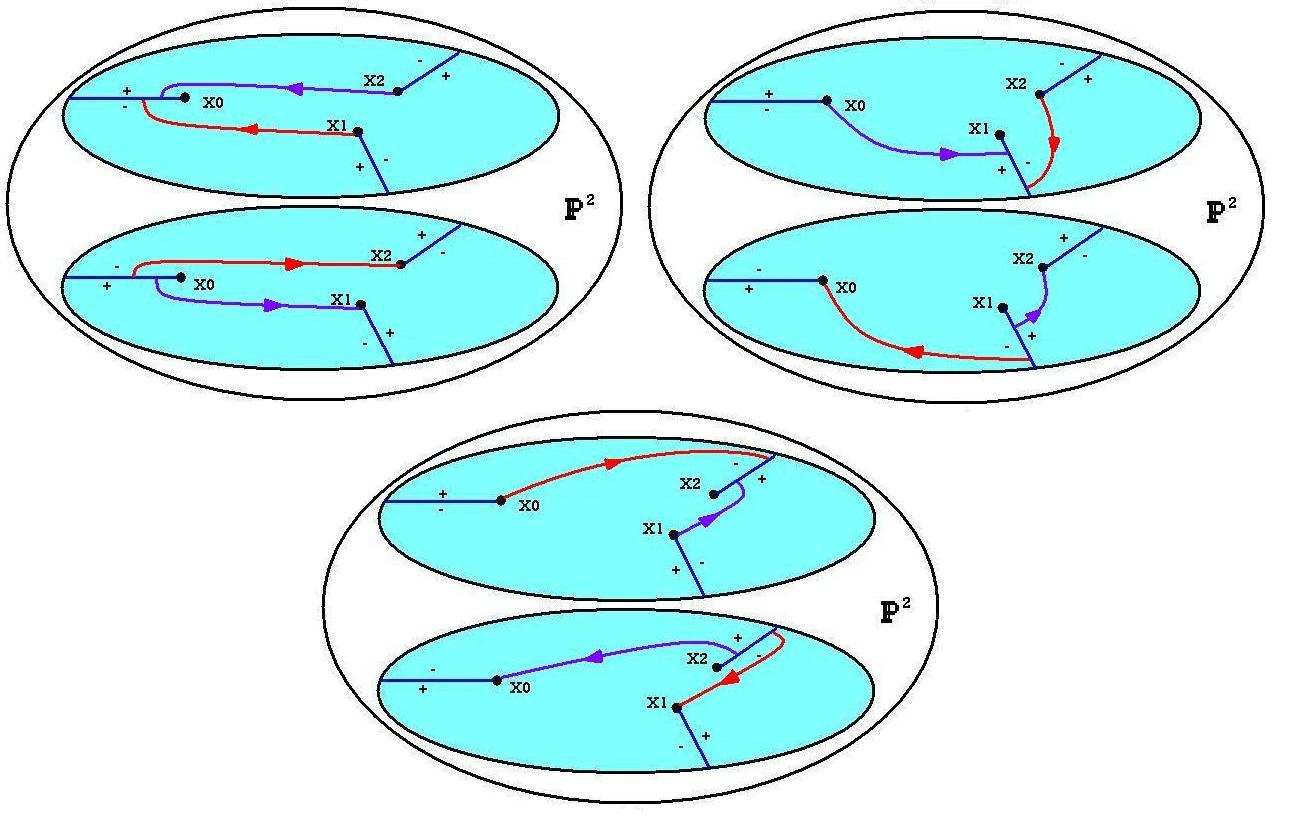}
\caption{Starting from the left, in clockwise order, the curves $\gamma_0, \gamma_1$ and $\gamma_2$, which collapse when $z$ goes to
$z_0, z_1$ and $z_2$ respectively.}\label{gamma-curve}
\end{figure}

\noindent Notice that in $\Sigma_*$ the three curves, as elements of $H_1(\Sigma_*)$, are not independent as they
satisfy the relation $[c_0]-[c_1]+[c_2]=0$. However, these are used to construct the thimbles in the elliptic fibration
\begin{eqnarray}
{D}_i =\left\{ (z_* + (z_i-z_*)t) \times c_i(z_* + (z_i-z_*)t)| t\in [0,1] \right\},
\end{eqnarray}
which are independent elements of $H_3(M)$. These Lagrangian cycles will be used as generators of the  Fukaya category of the mirror of $\PP^2$.\\
For our computations, it is convenient to introduce a more symmetric representation of the paths just described. Let us deform the cut $[X_1, X_2]$
so that it becomes the union of two paths from $X_i$ to $\infty$, $i=1,2$. Then we can choose a basis (in the above meaning of thimble
generators) of vanishing cycles $\gamma_j$, $j=0,1,2$, as shown on F{\sc igure} \ref{gamma-curve}. A direct inspection shows that
then $\gamma_j=(-1)^j c_j$.

\noindent Using the construction above these give us three 3-cycles ${\overline L}_i$ associated to the thimbles ${D}_i$.
The same cycles can be obtained as a double suspension as described at the end of section 2.

\subsection{Computation of the Periods}
Let us compute the periods of $\Omega(Y_y)$ along the Lagrangian cycles ${\overline L}_i$.\\
This means that we have to compute the integrals
\begin{eqnarray}
I_i=\frac{1}{(2\pi i)^3}\int_{{\overline L}_i} \frac {dx_1 \wedge dx_2 \wedge du}{x_1 x_2 u}=\frac{1}{(2\pi i)^2}\int_{{D}_i} \frac {dx_1 \wedge dx_2}{x_1 x_2},
\end{eqnarray}
where ${D}_i$ is the thimble generated by the vanishing cycle from $z_*$ to $z_i$.\\
The Lagrangian 3-cycles are better described in terms of the coordinates $(z,X,u)$. The restriction of $\Omega_3$ to the cycles is then easily obtained
by using (\ref{change}) in (\ref{holom3}) with $U=1$. This gives
\begin{eqnarray}
\Omega_3=\frac 12 dz\wedge \frac {dX}Y \wedge \frac {du}u.
\end{eqnarray}
Therefore, the periods are
\begin{eqnarray}
I_i=-\frac 1{8\pi^2} \int_{z_*}^{z_i} dz \int_{c_i} \frac {dX}{\sqrt{(X-X_0)(X-X_1)(X-X_2)}},
\end{eqnarray}
where the $X_i$'s are functions of $z$.
Let us choose to order the roots so that $X_j=X_k$ when $z=z_i$, $|\epsilon_{ijk}|=1$. Thus
\begin{eqnarray}
&& J_k:=\int_{c_k} \frac {dX}{\sqrt{(X-X_i)(X-X_j)(X-X_k)}}\cr
&& \phantom{J_i:}=2\int_{X_i}^{X_k} \frac {dX}{\sqrt{(X-X_i)(X-X_j)(X-X_k)}}-2\int_{X_k}^{X_j} \frac {dX}{\sqrt{(X-X_i)(X-X_j)(X-X_k)}}\cr
&& \phantom{J_i:}=\frac {2\pi}{\sqrt {X_j-X_i}}\ _2F_1\left(\frac 12, \frac 12; 1; \frac {X_k-X_i}{X_j-X_i}\right)
-\frac {2\pi}{\sqrt {X_i-X_j}}\ _2F_1\left(\frac 12, \frac 12; 1; \frac {X_k-X_j}{X_i-X_j}\right).
\end{eqnarray}
where $\ _2F_1$ is the hypergeometric function.\\
Next we have to integrate over $z$. To this end we can deform the path so that it passes through $z=0$ so that
\begin{eqnarray}
&& I_k=-\frac 1{8\pi^2}\int_{z_*}^{z_k} J_k dz= -\frac 1{8\pi^2}\int_0^{z_k} J_k dz
+\frac 1{4\pi}\int_0^{-y^{-\frac 13}} \frac {1}{\sqrt {X_j-X_i}}\ _2F_1\left (\frac 12, \frac 12; 1; \frac {X_k-X_i}{X_j-X_i}\right)dz\cr
&&\phantom{I_k=}-\frac 1{4\pi}\int_0^{-y^{-\frac 13}} \frac {1}{\sqrt {X_i-X_j}}\ _2F_1\left(\frac 12, \frac 12; 1; \frac {X_k-X_j}{X_i-X_j}\right)dz
=:A_k+B_k(y).
\end{eqnarray}
where we have used that $z_* = -y^{-\frac 1{3}}$.\\
The first integral in the last sum, $A_k$, does not depend on $y$. In order to compute $A_k$ it is convenient to rewrite these integrals
in terms of the paths $\gamma_k$ defined above. The paths $\gamma_k$ are symmetric
under the rotation given by multiplication by $\omega$. This is also true for the integration paths over $z$. Thus, as
the integrand is invariant under the same rotation and noticing that $c_k\simeq (-1)^k \gamma_k$, $k=0,1,2$, we see that
$A_k=(-1)^k A_0$. To compute $A_0$ we only need to observe that the 3-cycle associated to the sum $\gamma_0+\gamma_1+\gamma_2$
is homologically equivalent to the fundamental torus $T_0$ defined in \cite{h}, appendix A-2. This follows by noticing that the
cycle $[\Sigma_0 - D_0+ D_1- D_2]$ is the generator of $H^2(\CC^{*2},\ZZ)\simeq \ZZ$, see \cite{ako}, section 3.4, and
\cite{ako1}, proof of Lemma 4.9. As
\begin{eqnarray}
\frac{1}{(2\pi i)^3}\int_{T_0} \frac {dx_1 dx_2 du}{x_1 x_2 u}=1,
\end{eqnarray}
we get $A_0=\frac 13$, and then
\begin{eqnarray}
(A_0, A_1, A_2)=\left(\frac 13, -\frac 13, \frac 13\right).
\end{eqnarray}
Next we compute $B_k(y)$ up to order two in $y^{-\frac 13}$.
By setting
\begin{eqnarray}
f(z):=\frac 1{4\pi} \frac {1}{\sqrt {X_k-X_i}}\ _2F_1\left(\frac 12, \frac 12; 1; \frac {X_j-X_i}{X_k-X_i}\right),
\end{eqnarray}
we have:
\begin{eqnarray}
B_k(y)=-y^{-\frac 13} f(0)\ (-1)^k (\omega^{k+1}-\omega^{k-1})+ \frac 12 y^{-\frac 23} \frac {df}{dz}(0)\
(-1)^k (\omega^{k+1}-\omega^{k+2})+o(y^{-1}).
\end{eqnarray}
Using (\ref{radici}) we get
\begin{eqnarray}
f(0)=\frac {2^{\frac 13}}{4\pi \omega \sqrt {\omega^2-\omega}}\ _2F_1\left(\frac 12, \frac 12; 1; -\omega\right),
\end{eqnarray}
and from
\begin{eqnarray}
\frac d{dz} X^i (z=0)=\frac {\omega^{2i}}{2^{\frac 13} 3},
\end{eqnarray}
and the results in appendix \ref{app:calcoli} we finally have
\begin{eqnarray}
B_k(y)= \omega^k \frac {\sqrt 3}{8\pi^3} \Gamma(1/3)^3 y^{-\frac 13}+\omega^{2k} \frac {\sqrt 3}{16\pi^3} \Gamma(2/3)^3 y^{-\frac 23}+o(y^{-1}).
\end{eqnarray}
Thus,
\begin{eqnarray}\label{periods}
I_k(y)=\frac 13 (-1)^k +\omega^k \frac {\sqrt 3}{8\pi^3} \Gamma(1/3)^3 y^{-\frac 13}+\omega^{2k} \frac {\sqrt 3}{16\pi^3} \Gamma(2/3)^3 y^{-\frac 23}+o(y^{-1}).
\end{eqnarray}

\subsection{The Mirror Map}\label{mirrormap}
In order to determine the sheaves which are mirror to the Lagrangian cycles $L_i$ we need to relate the periods $I_k(y)$ to the hypergeometric functions $w_k(y)$.
To this end notice that both the $I_k$ and $w_k$ are solutions
of the same Picard-Fuchs system, so that we only need to compare them asymptotically for $y\to\infty$. To do this we start by computing the $w_i(y)$'s near $y=0$
and from this we will find their analytic extensions near $y = \infty$. Their expansions  near $y = 0$ are:
\begin{eqnarray}
&& w_0(y)= 1, \\
&& w_1(y)= \frac 1{2\pi i}\ln (y)+ \frac 3{2\pi i}\sum_{m=1} \frac {(3m-1)!}{(m!)^3} (-y)^m, \label{w1}\\
&& w_2(y)=-\frac 1{8\pi^2} (\ln (-y))^2 +\frac 18 -\frac {3}{4\pi^2} \ln (-y)\sum_{m=1} \frac {(3m-1)!}{(m!)^3} (-y)^m\cr
&& \phantom{w_2(y)=}-\frac {9}{4\pi^2} \sum_{n=1} \frac {(3n-1)!}{(n!)^3} [\psi(3n)-\psi(n+1)] (-y)^n.\label{w2}
\end{eqnarray}
Notice that we can write
\begin{eqnarray*}
&& \sum_{m=1} \frac {(3m-1)!}{(m!)^3} (-y)^m=\frac 1{2\pi i}\int_{-\frac 12-i\infty}^{-\frac 12+i\infty}
\frac {\Gamma(-3s)\Gamma(s)}{\Gamma(1-s)^2}y^{-s} ds,\\
&& \sum_{m=1} \frac {(3m-1)!}{(m!)^3} [\psi(3m)-\psi(m+1)] (-y)^m =\frac 1{2\pi i}\int_{-\frac 12-i\infty}^{-\frac 12+i\infty}
\frac {\Gamma(-3s)\Gamma(s)}{\Gamma(1-s)^2} [\psi(-3s)-\psi(1-s)] y^{-s} ds,
\end{eqnarray*}
which are easily verified by closing the path counterclockwise. The analytic extension at $y=\infty$ is obtained by closing the path clockwise.
A straightforward but tedious computation gives
\begin{eqnarray}
&& w_1(y)=\frac {3}{2\pi i} \left[ -\frac {y^{-\frac 13}}{4\pi^2} \sum_{n=0}^\infty \Gamma(n+1/3)^3 \frac {(-1)^n}{(3n+1)!} \frac {1}{y^n}
+\frac {y^{-\frac 23}}{4\pi^2} \sum_{n=0}^\infty \Gamma(n+2/3)^3 \frac {(-1)^n}{(3n+2)!} \frac {1}{y^n} \right]\\
&& w_2(y)=\frac 13+\frac {\sqrt 3}{4\pi} \left[ -(1+i\sqrt 3) \frac {y^{-\frac 13}}{4\pi^2} \sum_{n=0}^\infty \Gamma(n+1/3)^3 \frac {(-1)^n}{(3n+1)!}
\frac {1}{y^n} \right.\cr
&&\phantom{ w_2(y)=\frac 13+\frac {\sqrt 3}{4\pi}} \left.+(-1+i\sqrt 3)\frac {y^{-\frac 23}}{4\pi^2} \sum_{n=0}^\infty \Gamma(n+2/3)^3
\frac {(-1)^n}{(3n+2)!} \frac {1}{y^n} \right]
\end{eqnarray}
so that up to order two in $y^{-{\frac 1{3}}}$ we can write
\begin{eqnarray}\label{ws}
\begin{array}{l}
 w_0(y)=1\\
\\
 w_1(y)=-\frac {3}{8\pi^3 i} \Gamma(1/3)^3 y^{-\frac 13} +\frac 12 \frac {3}{8\pi^3 i} \Gamma(2/3)^3 y^{-\frac 23} +o(1/y)\\
\\
 w_2(y)=\frac 13 -\frac {\sqrt 3}{16\pi^3} (1+i\sqrt 3) \Gamma(1/3)^3 y^{-\frac 13} -
\frac 12 \frac {\sqrt 3}{16\pi^3} (1-i\sqrt 3) \Gamma(2/3)^3 y^{-\frac 23}+o(1/y)
\end{array}
\end{eqnarray}
By comparing the asymptotic expression of the periods (\ref{periods}) and (\ref{ws}) we get
\begin{eqnarray}
(w_0, w_1, w_2)= (I_0,I_1, I_2) \left(
\begin{array}{rrr}
1 & 0 & 0 \\ -1 & 1 & -1 \\ 1 & 1 & 0
\end{array}\right).\label{trasformazione}
\end{eqnarray}
For later convenience let us introduce the cycles
\begin{eqnarray}
\left(L^H_0, L^H_1, L^H_2\right) :=\left(-{\overline L}_1, -{\overline L}_0, {\overline L}_2\right),
\end{eqnarray}
where the minus sign means inversion of orientation. From (\ref{trasformazione}) we see that if we define
\begin{eqnarray}
\begin{array}{l}
N_0 := {\mathcal B}_2  = [\scro_{\bp}(-2)], \\
\\
N_1 :=- {\mathcal B}_0 + {\mathcal B}_1 +2 {\mathcal B}_2 = -[\scro_p] + [\scro_H (-1)] + 2 [\scro_{\bp}(-2)], \\
\\
N_2 := {\mathcal B}_1 + {\mathcal B}_2 = [\scro_H (-1)] + [\scro_{\bp}(-2)]
\end{array}
\end{eqnarray}
the physical mirror map on objects gives:
\begin{eqnarray}
&& {Mir}^H: \left({ {N}}_0, { {N}}_1, { {N}}_2\right)\mapsto \left(L^H_0, L^H_1, L^H_2\right). \label{rimappa}
\end{eqnarray}
On the other hand, as we will see in the next section, from \cite{ako} we know that ${\rm Hom}\left( N_i, N_j \right) \cong {\rm Hom} \left( L_i, L_j \right)$;
then using Seidel's results (proposition 4.4 and 5.1 in \cite{Seidel1}) one can define the functor on morphisms. We can then state our main result in the
following proposition
\begin{prop}
Let $X= Tot(K_{\PP^2})$ and $Y$ be its mirror. Denote by $D^b_{\PP^2} \bigl(Coh(X)$ the derived category of the coherent sheaves of the total space of the
canonical bundle of $\PP^2$ supported on the zero section. Then the functor
\begin{eqnarray}
 {Mir}^{H}: D^b_{\PP^2} \bigl(Coh(X) \bigr) \rightarrow {Lag}_{vc}(Y),
\ \left({ {N}}_0, { {N}}_1, { {N}}_2\right)\mapsto \left(L^H_0, L^H_1, L^H_2\right),
\end{eqnarray}
defines an equivalence between $D^b_{\PP^2} \bigl(Coh(X)$ and the full subcategory ${Lag}_{vc}(Y)$, generated by the cycles $L^H_i$, $i=1,2,3$,
of the derived category associated to the $A_\infty$ Fukaya category of vanishing Lagrangian cycles in the hypersurface $Y\subset \CC^2\times \CC^{*2}$.
\end{prop}

\subsection{Discussion}
As mentioned in section 2, the equivalence between these two categories has been established by P. Seidel in \cite{Seidel1},
extending to toric del Pezzo surfaces the equivalence of \cite{ako} by means of a double suspension.
However, the functor realizing the equivalence is not the same as the one we have just constructed by using the physical prescription suggested by Hosono.
Indeed, the mirror map in \cite{ako}, lifted to the total space of the canonical bundle of $\PP^2$, associates to the cycles $ L_i$ the sheaves $ E_i$, where
\begin{eqnarray}
&& \pmb {\tau}\equiv \left\{ {E}_0, {E}_1, {E}_2\right\}:=\left\{ {\mathcal O}, {\mathcal T}(-1),
\Lambda^2 {\mathcal T}(-2)\simeq {\mathcal O}(1)\right\}.
\end{eqnarray}
The direct image of the $ E_i$'s under $\iota: \PP^2 \to X$ generates $D^b_{\PP^2} \bigl(Coh(X) \bigr)$. In K-theory these are related to the objects in
$\pmb \omega= \Bigl\{ [{ {\mathcal O}}_p],[{ {\mathcal O}}_H (-1)],[{ {\mathcal O}}_{\PP^2}(-2)] \Bigr\}$ by
\begin{eqnarray}
\pmb \omega =\pmb \tau C_{\tau\omega}, \qquad C_{\tau \omega}=
\left(\begin{array}{rrr}
1 & -3 & 6 \\ -1 & 2 & -3 \\ 1 & -1 & 1
\end{array}\right).                           \label{newbasis}
\end{eqnarray}
The transformation matrix in (\ref{newbasis}) is not the same as the one in (\ref{trasformazione}),
therefore the equivalence defined in \cite{ako} does not provide
the physical mirror map realizing the physical correspondence defined by the central charge, see \cite{h}.\\
Let us take a closer look at the objects $ N_i$. By abuse of notation we use the same symbol both for a sheaf $F$ on $\PP^2$ and its direct image $\iota_* F$
on $X$.
Up to now the $ N_i$ have been elements in K-theory. We will show that they can be associated to exceptional sheaves on $\PP^2$ and relate them to the ones used
in \cite{ako}.\\
Since
\begin{eqnarray}
\begin{array}{c}
\left[ \scro_p \right] = \left[ \scro_{\bp} \right] -2 \left[ \scro_{\bp}(-1) \right] + \left[ \scro_{\bp}(-2) \right]\\
\\
\left[ \scro_H(-1) \right] = \left[ \scro_{\bp}(-1) \right] - \left[ \scro_{\bp}(-2) \right]
\end{array}
\end{eqnarray}
we can write the K-classes of the $N_i$ as
\begin{eqnarray}
\begin{array}{l}
N_0 = [\scrop(-2)],\\
\\
N_1 = -[\scrop] + 3[\scrop(-1)],\\
\\
N_2 = [\scrop(-1)].
\end{array}
\end{eqnarray}
By tensoring the Euler sequence for $\bp$ by $\scro(-1)$ we get that at the level of K-theory $N_1 = {\mathcal T}(-3)$.\\
Thus we can assume
\begin{eqnarray}
\begin{array}{l}
N_0 = \scrop(-2),\\
\\
N_1 = {\mathcal T}(-3),\\
\\
N_2 = \scrop(-1)
\end{array}
\end{eqnarray}
as objects in $D^b (Coh(\PP^2))$ (or $D^b_{\PP^2} \bigl(Coh(X) \bigr)$ after taking the direct image of both sides).\\
Since this is a strong exceptional collection for $\bp$, its direct image in $D^b(Coh(X))$ generates $D^b_{\PP^2}(Coh(X))$. This exceptional sequence is related
to the one used in \cite{ako} by an autoequivalence which is tensoring by $\scro(-2)$.
One can compute the morphisms between the sheaves by standard methods. If we write $\bp=\PP(V)$, for $V\cong \CC^3$ , then:
\begin{eqnarray}
\begin{array}{ll}
 {\rm Hom}^i_{\PP^2}\left( N_0,  N_1\right) \cong {\rm Hom}^i_{\PP^2}\left(\scrop(-2), {\mathcal T}(-3)\right) \cong \wedge^2 V^* & \text{for }  i= 0, \\
\\
 {\rm Hom}^i_{\PP^2}\left( N_0, N_2\right) \cong {\rm Hom}^i_{\PP^2}\left(\scrop(-2), \scrop(-1)\right) \cong V^* & \text{for }  i= 0,\\
\\
{\rm Hom}^i_{\PP^2}\left(N_1, N_2\right) \cong {\rm Hom}^i_{\PP^2}\left({\mathcal T}(-3), \scrop(-1)\right) \cong V & \text{for } i=0,
\end{array}
\end{eqnarray}
and all the other homomorphism groups are zero. For the direct image in $X$ we see that
${\rm Hom}^0_{X}\left(\iota_* N_0, \iota_*  N_1\right) = \wedge^2 V^* = {\rm Hom}^3_{X}\left(\iota_* N_1, \iota_*  N_0\right)^*$ and zero otherwise. Similarly for
the other morphisms. These are all three dimensional agreeing with the fact that $| L^H_i \cap  L^H_j|=3$ for $i\neq j$.

\section{Conclusions}
We have considered the mirror map relating the derived category of coherent sheaves on $X=K_{{\mathbb{P}}^2}$ (with cohomology supported on the zero section)
and the Fukaya category of Lagrangian vanishing 3-cycles on its mirror $Y$. \\
For the projective plane the mirror map has been studied by Aroux, Katzarkov and Orlov in \cite{ako} and Seidel \cite{Seidel3}. In that case the mirror
equivalence was realized by associating to an exceptional collection of sheaves on ${\mathbb{P}}^2$, an exceptional collection of thimbles on the mirror,
corresponding to suitable vanishing $S^1$-cycles on a smooth elliptic fiber of the mirror manifold. The extension of the mirror map to the total space of the
canonical bundle $K_S$ of a toric del Pezzo surface was studied by Seidel in \cite{Seidel1}. An exceptional collection of sheaves on ${\mathbb{P}}^2$ is
extended to a collection of spherical objects on $X= K_{\PP^2}$, generating the category $D^{b}_{ {\mathbb{P}}^2}\bigl(Coh(X) \bigr)$. On the mirror of $X$, $Y$, the
thimbles considered in \cite{ako} are promoted to Lagrangian 3-cycles by a double suspension of the underlying $S^1$-cycles.
The functor obtained in this way realizes an equivalence between $D^{b}_{{\mathbb{P}}^2}\bigl(Coh(X) \bigr)$ and a full subcategory of ${{Lag}}^0_{vc}(Y)$.\\
In the present paper we have shown that the map constructed this way does not coincide with the one expected following the physical prescription, in fact the periods of the chosen Lagrangian 3-cycles of $Y$ do not reproduce the correct central charges of the corresponding B-branes in $X$ represented by the
mirror coherent sheaves. The physical map can be obtained by composing the one in \cite{Seidel1} with an autoequivalence
of categories. We have determined the autoequivalence, and thus the physical mirror map, by performing an explicit computation
of the periods of the Lagrangian 3-cycles and comparing them with the central charges. As a byproduct we have obtained a check, for
$X=K_{{\mathbb{P}}^2}$, of the conjecture stated in \cite{h} for noncompact threefolds.\\
By construction, our physical mirror map is compatible with the mirror map identifying the K\"ahler and the complex moduli spaces used for the physical computation of Gromov-Witten invariants. This fact could be useful in order to gain a better and complete understanding of the various aspects of mirror symmetry. For example, our map could help the analysis of the relation between monodromy, GW invariants and wall crossing. Moreover, it will be interesting to extend our analysis to the total space of the canonical bundle of del Pezzo surfaces and, in particular, to the (resolution of) non toric orbifold $\mathbb{C}^3/\Delta_{27}$ \cite{Cacciatori:2010ci}.

\vspace{3cm}
\section*{Acknowledgments}
\emph{We are grateful to Bert van Geemen for useful discussions. We also thank Peter Dalakov for providing helpful comments on the manuscript.}

\newpage
\begin{appendix}

\section{Evaluation of the hypergeometric functions}\label{app:calcoli}
Here we will compute the quantities
\begin{eqnarray}
&& F(-\omega):=\ _2F_1 \left(\frac 12, \frac 12;1; -\omega\right),\\
&& F'(-\omega):=\ _2F_1' \left(\frac 12, \frac 12;1; -\omega\right),
\end{eqnarray}
where the prime means derivation w.r.t. the argument. The starting point is a result due to Ramanujan (\cite{Rama}, Chapter 11, Entry 32(v)):
\begin{eqnarray}\label{ramanujan}
(1+x^2)^{\frac 14} F\left(\frac 12 (1+ix)\right)=\frac {1+i}2 F\left(\frac 12 \left( 1+\frac x{\sqrt {1+x^2}} \right)\right)+
\frac {1-i}2 F\left(\frac 12 \left( 1-\frac x{\sqrt {1+x^2}} \right)\right).
\end{eqnarray}
We can specialize this formula to $x=\sqrt 3$:
\begin{eqnarray}
F(-\omega)=\frac {1+i}{2\sqrt 2} F\left(\frac 12 \left( 1+\frac {\sqrt 3}2 \right)\right)+
\frac {1-i}{2\sqrt 2} F\left(\frac 12 \left( 1-\frac {\sqrt 3}2 \right)\right).
\end{eqnarray}
Now use
\begin{eqnarray}
\frac 12 \left( 1\pm\frac {\sqrt 3}2 \right)=\left(\frac {\sqrt 6 \pm \sqrt 2}{4} \right)^2=:k_\pm,
\end{eqnarray}
and
\begin{eqnarray}
F(k^2)=\frac 2\pi K(k),
\end{eqnarray}
where $K$ is the complete elliptic integral of the first kind, to get
\begin{eqnarray}
F(-\omega)=\frac {1+i}{\pi\sqrt 2} K(k_+)+\frac {1-i}{\pi\sqrt 2} K(k_-).
\end{eqnarray}
Finally, we note that $k_\pm$ are two of the so called {\it singular values} for the elliptic integrals, so that $K(k_\pm)$
are explicitly calculable \cite{SelCho}:
\begin{eqnarray}
K(k_+)=\frac 1\pi 2^{-\frac 73} 3^{\frac 34} \Gamma(1/3)^3, \qquad K(k_-)=\frac 1\pi 2^{-\frac 73} 3^{\frac 14} \Gamma(1/3)^3.
\end{eqnarray}
Then,
\begin{eqnarray}
F(-\omega)=\frac {1+i}{\pi^2\sqrt 2} 2^{-\frac 73} 3^{\frac 34} \Gamma(1/3)^3 +\frac {1-i}{\pi^2\sqrt 2} 2^{-\frac 73} 3^{\frac 14} \Gamma(1/3)^3.
\end{eqnarray}
Now we proceed in computing the first derivative. Define
\begin{eqnarray}
y_\pm^2:=x_\pm:=1\pm\frac x{\sqrt {1+x^2}},
\end{eqnarray}
end differentiate (\ref{ramanujan}) w.r.t. x:
\begin{eqnarray}
&& \frac {x}{2(1+x^2)^{\frac 34}} F((1+ix)/2)+\frac i2 (1+x^2)^{\frac 14} F'((1+ix)/2)=
\frac {1+i}2 \frac {dx_+}{dx} F'(x_+) +\frac {1-i}2 \frac {dx_-}{dx} F'(x_-)\cr
&& \qquad\ =\frac {1+i}{4y_+} \frac {dx_+}{dx} \frac {d}{dy_+} F(y_+^2) +\frac {1-i}{4y_-} \frac {dx_-}{dx} \frac {d}{dy_-} F(y_-^2)\cr
&& \qquad\ =\frac {1+i}{2\pi y_+} \frac {dx_+}{dx} K'(y_+) +\frac {1-i}{4y_-} \frac {dx_-}{dx} K'(y_-) \cr
&& \qquad\ =\frac {1+i}{2\pi} \frac {dx_+}{dx} \left( \frac {E(y_+)}{x_+(1-x_+)} -\frac {K(y_+)}{x_+}\right)+
\frac {1-i}{2\pi} \frac {dx_-}{dx} \left( \frac {E(y_-)}{x_-(1-x_-)} -\frac {K(y_-)}{x_-}\right),
\end{eqnarray}
where $E$ is the second elliptic integral related to the first one by
\begin{eqnarray}
K'(k)=\frac {E(k)}{k(1-k^2)} -\frac {K(k)}{k}.
\end{eqnarray}
Using the previous results, $k_\pm =y_\pm(\sqrt 3)$, and
\begin{eqnarray}
&& E(k_+)=2^{\frac 13} 3^{-\frac 14} \pi^2 \Gamma (1/3)^{-3}+2^{-\frac{10}3} 3^{\frac 14} \pi^{-1} (\sqrt 3-1)\Gamma(1/3)^3,\\
&& E(k_-)=2^{\frac 13} 3^{-\frac 34} \pi^2 \Gamma (1/3)^{-3}+2^{-\frac{10}3} 3^{-\frac 14} \pi^{-1} (\sqrt 3+1)\Gamma(1/3)^3,
\end{eqnarray}
we get
\begin{eqnarray}
F'(-\omega)=\frac 1{2\pi^2} \Gamma(1/3)^3 3^{-\frac 14} 2^{-\frac 73} (i^{\frac 12} -\sqrt 3 i^{-\frac 12})
+\frac 1{\pi^2} \Gamma(2/3)^3 3^{\frac 34} 2^{-\frac 83} (i^{\frac 12} +\sqrt 3 i^{-\frac 12}).
\end{eqnarray}

\end{appendix}

\newpage



\begin{thebibliography}{99}

\bibitem{ako}
   D.~ Auroux, L.~Katzarkov, D.~Orlov,
   ``Mirror symmetry for del Pezzo surfaces: vanishing cycles and
     coherent sheaves'',
   Invent. Math.  166  (2006), 537--582.

\bibitem{ako1}
   D.~ Auroux, L.~Katzarkov, D.~Orlov,
   ``Mirror symmetry for weigthed projective planes and their noncommutative deformations'',
   Ann. of Math. (2) 167 (2008), no. 3, 867–943.

\bibitem{Ballard}
   M.~Ballard,
   ``Sheaves on local Calabi-Yau varieties,''
   arXiv:0801.3499.

\bibitem{Rama}
B.~C.~Berndt,
``Ramanujan's Notebooks, Part II,'' Springer Verlag (1989) New York.

\bibitem{Bouchard:2007nr}
  V.~Bouchard and R.~Cavalieri,
  ``On the mathematics and physics of high genus invariants of $[C^3/Z_3]$'',
  Adv.\ Theor.\ Math.\ Phys.\  {\bf 13} (2009) 695.

\bibitem{Cacciatori:2008fq}
  S.~L.~Cacciatori and M.~Compagnoni,
  ``D-Branes on $C^3_6$ part I: prepotential and GW-invariants'',
  Adv.\ Theor.\ Math.\ Phys.\  {\bf 13}, 1371 (2009).

\bibitem{Cacciatori:2010ci}
  S.~L.~Cacciatori and M.~Compagnoni,
  ``On the geometry of $C^3/\Delta_{27}$ and del Pezzo surfaces,''  JHEP {\bf 1005} (2010) 078.

\bibitem{SelCho}
   S. Chowla and A. Selberg, "On Epstein's Zeta-Function." J. reine angew. Math. 227, 86-110, 1967

\bibitem{Chiang}
   T.-T. Chiang, A. Klemm, S.-T. Yau and E. Zaslow,
   ``Local mirror symmetry: calculations and interpretations'',
   Adv. Theor. Math. Phys. \textbf{3} (1999), 495.

\bibitem{Craw}
   A. Craw,
   ``An explicit construction of the McKay correspondence for A-Hilb",
   J. Algebra {\bf 285}, (2005), 682--705.

\bibitem{ossa}
   X.~De la Ossa, B.~Florea and H.~Skarke,
   ``D-branes on noncompact Calabi--Yau manifolds: K-theory and monodromy",
   Nucl.\ Phys.\  B {\bf 644} (2002), 170.

\bibitem{FultonT}
   W. ~Fulton,
   ``Introduction to toric varieties",
   Annals of Mathematics Studies, \textbf{131},
   Princeton University Press, Princeton, NJ, 1993.

\bibitem{GKZ}
   I. M. Gelfand, A. V. Zelevinski and M. M. Kapranov,
   {Equations of hypergeometric type and toric varieties},
   \textit{Funktsional Anal. i. Prilozhen.} \textbf{23} (1989), 12--26; English transl.
   Functional Anal. Appl. \textbf{23} (1989),\break 94--106.

\bibitem{hiv}
  K.~Hori, A.~Iqbal and C.~Vafa,
  ``D-branes and mirror symmetry,''
  arXiv:hep-th/0005247.

\bibitem{h}
  S.~Hosono,
  ``Central charges, symplectic forms, and hypergeometric series in local mirror symmetry'', Mirror symmetry. V,
  405--439, AMS/IP Stud. Adv. Math., 38, Amer. Math. Soc., Providence, RI, 2006.

\bibitem{HKTY1}
   S.~Hosono, A.~Klemm, S.~Theisen and S.~T.~Yau,
   \textit{Mirror symmetry, mirror map and applications to Calabi--Yau
     hypersurfaces},
   Commun.\ Math.\ Phys.\  {\bf 167} (1995), 301.

\bibitem{HKTY2}
   S.~Hosono, A.~Klemm, S.~Theisen and S.~T.~Yau,
   \textit{Mirror symmetry, mirror map and applications to complete intersection
     Calabi--Yau spaces},
   Nucl.\ Phys.\  B {\bf 433} (1995), 501.

\bibitem{HLY1}
   S.~Hosono, B.~H.~Lian and S.~T.~Yau,
   \textit{GKZ generalized hypergeometric systems in mirror symmetry of Calabi--Yau
     hypersurfaces},
   Commun.\ Math.\ Phys.\  {\bf 182} (1996), 535.

\bibitem{HLY2}
   S.~Hosono, B.~H.~Lian and S.~T.~Yau,
   \textit{Maximal degeneracy points of GKZ systems},
   arXiv:alg-geom/9603014.

\bibitem{Ito-Nakajima}
   Y.~Ito and H.~Nakajima,
 ``McKay correspondence and Hilbert schemes in dimension three",
  Topology \textbf{39} (2000), 1155--1191.

\bibitem{Karp:2006pb}
  R.~L.~Karp,
 ``On the $C^n/Z_m$ fractional brane``,
  J.\ Math.\ Phys.\  {\bf 50}, 022304 (2009).

\bibitem{kontsevich}
   M.~Kontsevich,
  ``Homological algebra of mirror symmetry",
   alg-geom/9411018.

\bibitem{Segal}
   E.~Segal,
   ``The $A_\infty$ deformation theory of a point and the derived category of local Calabi-Yaus,''
   J.~Algebra, 320(8):3232-3268, 2008.

\bibitem{Seidel}
P.~Seidel,
``Vanishing cycles and mutations''
European Congress of Mathematics, Vol. II (Barcelona, 2000), 65–85, Progr. Math., 202, Birkhäuser, Basel, 2001.

\bibitem{Seidel3}
P.~Seidel,
``More about vanishing cycles and mutations''
Symplectic geometry and mirror symmetry ({S}eoul, 2000), 429--465, World Sci. Publ., River Edge, NJ, 2001

\bibitem{Seidel1}
P.~Seidel,
``Suspending Lefschetz fibrations, with an application to Local Mirror Symmetry,'' Comm.~Math.~Phys. 297 (2010) 515-528.

\bibitem{Seidel2}
P.~Seidel,
``Fukaya Categories and Picard--Lefschetz Theory,''
2008 European Mathematical Society, ISBN 978-3-03719-063-0.

\end{thebibliography}
\end{document}